\documentclass[12pt]{article}

\usepackage[utf8]{inputenc}
\usepackage[T1]{fontenc}
\usepackage{amsmath,amssymb}
\usepackage{bm}
\usepackage[left=0.85in,right=0.85in,top=1.0in,bottom=1.0in]{geometry}
\usepackage{graphicx}
\usepackage{siunitx}
\usepackage{setspace}
\usepackage{textcomp}
\usepackage{titlesec}
\usepackage{tocloft}
\usepackage[numbers,sort&compress]{natbib}
\usepackage[hidelinks]{hyperref}

\titleformat{\section}{\large\bfseries}{\thesection}{0.75em}{}
\titlespacing*{\section}{0pt}{1.8ex plus 0.4ex minus 0.2ex}{0.8ex}
\titleformat{\subsection}{\normalsize\bfseries}{\thesubsection}{0.75em}{}
\titlespacing*{\subsection}{0pt}{1.2ex plus 0.3ex minus 0.2ex}{0.5ex}

\begin{document}

\begin{center}
{\Large\bfseries Sensitive endoscopic diamond magnetometer for non-contact sensing in confined environments\par}
\vspace{1em}
Johannes Wesseler$^{1}$ and Roland Nagy$^{1,*}$\par
\vspace{0.5em}
$^{1}$Institute of Applied Quantum Technologies, Friedrich-Alexander-Universit\"at Erlangen-N\"urnberg, Erlangen, Germany\par
\vspace{0.5em}
$^*$Correspondence: roland.nagy@fau.de
\end{center}

\begin{abstract}

Transitioning quantum magnetometry from laboratory environments to real-world applications has been limited by a persistent trade-off between sensor miniaturization and magnetic sensitivity. While bulky systems can achieve high sensitivity, endoscopic probes commonly suffer from inefficient fluorescence collection and reduced performance. Here we resolve this trade-off and present a miniaturized diamond quantum magnetometer with a 6 mm diameter endoscopic sensor head, achieving a magnetic-field sensitivity of $91 \, \mathrm{pT}/\sqrt{\mathrm{Hz}}$ with a 2 kHz measurement bandwidth in a magnetically unshielded environment. The fluorescence collection bottleneck is overcome by separating excitation and collection into different cores of a fused multi-core fiber bundle, coupled to the diamond through a custom high-numerical-aperture micro-objective. A compact FPGA-based backend performs microwave control, lock-in detection and real-time resonance tracking, enabling robust operation during magnetic-field imaging. To demonstrate the practical utility of the miniaturized sensor, we image the magnetic field of a commercial lithium-ion pouch cell during charge and discharge and reconstruct depth-integrated current-density maps of the current flow. These results show that endoscopic diamond magnetometers can combine high sensitivity with a probe geometry suitable for confined, unshielded measurements, opening new avenues in battery technology and beyond.
\end{abstract}

\section*{Introduction}

Magnetic field measurements can provide a non-contact signature of electrical currents and magnetic materials. Imaging these fields with highly sensitive quantum magnetometers is therefore useful for applications ranging from biomedical measurements to non-destructive testing and \emph{in operando} diagnostics of technical systems \cite{lenz_magnetic_2006,ripka_magnetic_2021,wikswo_magnetic_1980,baillet_magnetoencephalography_2017,pointner_optimizing_2025,pointner_compact_2026,zamudio_ramirez_magnetic_2022,hollendonner_quantum_2023}. In many practical settings, however, the limiting requirement is not sensitivity alone. A useful sensor must combine high magnetic-field sensitivity with a probe head that can operate close to the object, tolerate spurious magnetic fields from nearby sources, respond quickly during scanning or dynamic operation, and fit into confined spaces.

Existing magnetometers cover only parts of this requirement space. Hall and fluxgate sensors are robust and compact, but their sensitivities typically cannot reach those of quantum sensors. Superconducting quantum interference devices (SQUIDs) and optically pumped atomic magnetometers (OPMs), for example, can reach sub-$\mathrm{fT}/\sqrt{\mathrm{Hz}}$ sensitivities \cite{storm_ultra-sensitive_2017,kominis_subfemtotesla_2003,dang_ultrahigh_2010}. However, they commonly require stringent temperature control (cryogenics for SQUIDs, heating for OPMs) and magnetic shielding, which increases package size and limits deployability. In contrast, quantum sensors based on nitrogen-vacancy (NV) centers in diamond operate under ambient conditions and combine high dynamic range, broad bandwidth and excellent magnetic-field sensitivity \cite{doherty_nitrogen_vacancy_2013,rondin_magnetometry_2014,barry_sensitivity_2020}. Ensemble NV magnetometers have reached sensitivities of a few hundred $\mathrm{fT}/\sqrt{\mathrm{Hz}}$ \cite{barry_sensitive_2024,fescenko_diamond_2020}, and recent work has moved these systems from experimental setups toward compact sensing devices \cite{graham_fiber_coupled_2023, wang_fully_2025}. The key challenge for endoscopic NV magnetometry is to retain this performance while reducing the sensor head to a size and geometry suitable for endoscopic access. Miniaturized NV magnetometers have so far followed two main design strategies. One approach integrates all components required for optical excitation, fluorescence collection and microwave excitation close to the diamond in a compact sensor package \cite{wang_fully_2025, kumar_high_2024}. This can provide efficient fluorescence collection and high sensitivity, but the sensing unit itself becomes physically larger and thus less easily deployable in confined spaces. A second approach keeps the laser, detector and control electronics in a remote backend and connects them to a small sensing probe through optical fibers and electrical wiring \cite{patel_subnanotesla_2020,graham_fiber_coupled_2023,chatzidrosos_fiberized_2021,hatano_simultaneous_2021,dix_fiber-tip_2022,homrighausen_microscale_2024,johansson_miniaturized_2025,newman_endoscopic_2025,singh_fiber_integrated_2026}. This architecture is well suited to endoscopic or otherwise hard-to-access measurements, but it introduces a physical bottleneck: NV fluorescence is emitted over a large optical etendue inside the diamond \cite{wolf_subpicotesla_2015}, whereas the fiber accepts only a small fraction of this light due to its limited etendue \cite{hodara_throughput_1984}. In single-fiber probes, increasing the fiber core diameter or numerical aperture can increase the fiber etendue, but it also tends to increase the excitation volume in the diamond and therefore the source etendue of the fluorescence. The effective gain in collection efficiency is therefore limited. Placing the detector directly in the sensor head can avoid part of this loss, but again increases probe size and complexity \cite{sturner_integrated_2021,xie_microfabricated_2022}.

Here we show for the first time an endoscopic NV magnetometer that addresses this collection bottleneck by separating excitation and collection in a fused multi-core fiber bundle. A small central fiber delivers the excitation light to keep the extent of the excited region limited, while four larger surrounding fibers collect fluorescence through a high-numerical-aperture micro-objective at the sensor tip. This miniaturized optical architecture is the enabling element of the sensor: it increases the collected fluorescence without requiring a detector or bulk optics at the probe head, thereby preserving the deployability of a 6 mm diameter endoscopic geometry. The sensor head is combined with a compact backend based on a field-programmable gate array (FPGA). The backend generates microwave control tones, performs digital lock-in detection and tracks the optically detected magnetic resonance (ODMR) frequency in real time. Closed-loop resonance tracking allows continuous magnetic-field measurements without repeated full-spectrum ODMR acquisitions, which is important for fast scans. It further enables measurements in which the magnetic field, ODMR contrast or linewidth changes during operation, e.g., during scans over metallic components. We characterize the sensor and obtain an open-loop magnetic-field sensitivity of $91 \, \mathrm{pT}/\sqrt{\mathrm{Hz}}$ over a measurement bandwidth of $2 \, \mathrm{kHz}$ in a magnetically unshielded environment.

We then use the sensor for magnetic-field imaging of a commercial lithium-ion pouch cell during charge and discharge. Magnetic imaging of current flow in batteries has been increasingly explored for non-destructive testing and \emph{in operando} diagnostics \cite{brauchle_defect_2023,lee_diagnosis_2023,suzuki_non-destructive_2021}. Optically pumped magnetometers have been applied to battery measurements, but these experiments typically require magnetic shielding and offer limited dynamic magnetic-field range \cite{hu_sensitive_2020, evans_quantum_2025}. This complicates their application in realistic environments such as assembly lines or with packaged cells that may have strong magnetic signatures from the tab materials and connectors.
NV-based scanning probe microscopy has also been proposed for battery magnetic-field mapping, especially for nanoscopic field features \cite{pollok_magnetic_2025}; however, the small scan area and slow scan speed limit its use for larger packaged cells. Our endoscopic sensor instead allows sensitive, high-throughput magnetic-field mapping over larger areas in unshielded environments, with the compact sensor head placed close to the object while the laser, detector and microwave electronics remain remote. From the measured magnetic fields we reconstruct the large-scale, depth-integrated sheet-current density inside the cell, demonstrating the ability of the platform to perform quantitative magnetic imaging in a practical measurement geometry.

\section*{Results and Discussion}

\subsection*{Endoscopic sensor architecture}

\begin{figure}[!htbp]
    \centering
    \includegraphics[width=\linewidth]{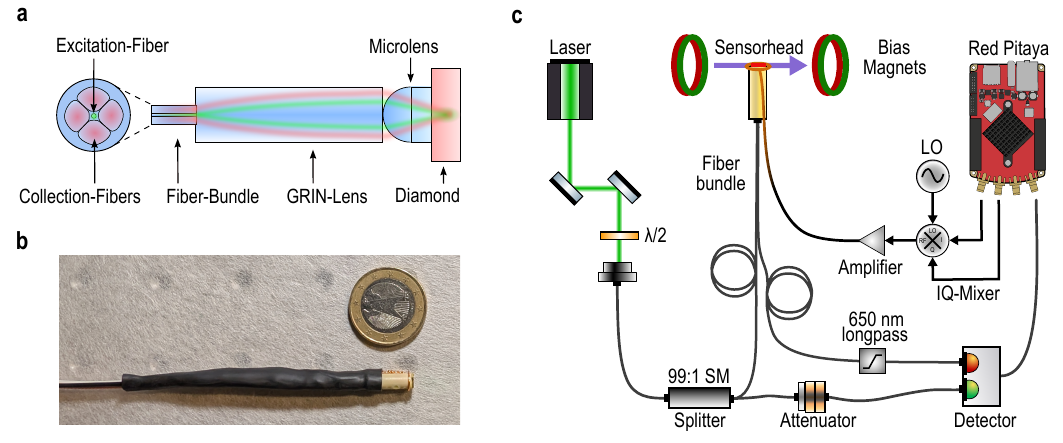}
    \caption{\textbf{Endoscopic NV magnetometer architecture.} \textbf{a}, Schematic of the endoscopic sensor head optics. Excitation light is delivered through the central fiber of the fused fiber bundle and focused into the diamond by the micro-objective, while NV fluorescence is collected through the four outer fibers. \textbf{b}, Photograph of the packaged 6 mm diameter sensor head next to a 1-euro coin for scale. \textbf{c}, Schematic of the complete measurement setup. The sensor receives microwave control signals from the FPGA-based backend, while the laser and photodetector are connected through the fiber bundle. Detection is performed with an auto-balanced photodetector for laser-noise cancellation.}
    \label{fig:sensorhead}
\end{figure}

The magnetometer is built around a 6 mm diameter sensor head that contains the diamond, a compact micro-objective and a fused multi-core fiber bundle (Fig.~\ref{fig:sensorhead}a,b). The design goal is to keep the probe head small enough for endoscopic deployment while preserving the fluorescence collection required for high magnetic sensitivity. To achieve this, the optical excitation path is separated from the fluorescence collection path. The central fiber has a 14 \textmu m core diameter and numerical aperture of 0.12, so the excitation light can be imaged into a small volume in the diamond. Four outer fibers, each with a 200 \textmu m core diameter and numerical aperture of 0.22, collect fluorescence and guide it back to the detector. Between the fiber bundle and the diamond, a 0.25-pitch gradient-index lens and a 1 mm diameter spherical lens form a high-numerical-aperture micro-objective. This optical layout follows a principle widely used in endomicroscopy: excitation and collection are separated and thus can be optimized independently while the probe remains fiber-based \cite{flusberg_fiber-optic_2005,beaudette_double-clad_2022, matz_design_2016}.

This separation is central to the sensor performance. In the shot-noise limit of continuous-wave ODMR magnetometry, \(\eta_{\mathrm{SN}}=(4/3\sqrt{3})\Delta\nu/(\gamma_e C\sqrt{R})\), where \(\Delta\nu\) is the ODMR linewidth, \(C\) is the contrast, \(\gamma_e\) is the electron gyromagnetic ratio and \(R\) is the detected photon rate \cite{dreau_avoiding_2011}. The fluorescence rate must therefore be increased without expanding the excitation volume so far that the source etendue again exceeds the acceptance of the collection fibers. In the present design, the small central fiber defines the excitation region, while the four larger surrounding fibers provide the collection etendue. Optical simulations give a lateral excitation full width at half maximum of 2.9 \textmu m and a depth of focus of 53 \textmu m, corresponding to an approximate excitation volume of $350 \, \mu\mathrm{m}^3$, and a fluorescence collection efficiency of 2.5\% (Supplementary Note 2). Because excitation and collection propagate in separate fiber cores, this geometry can also reduce background autofluorescence generated when high laser powers excite photoactive compounds in the optical fiber \cite{johansson_miniaturized_2025}. The same tight excitation volume improves the practical scan resolution compared with larger single-fiber designs \cite{newman_endoscopic_2025} and could be advantageous for pulsed protocols that require high optical intensities \cite{wang_comparing_2024}.

The remote backend includes the laser, balanced detector and microwave electronics (Fig.~\ref{fig:sensorhead}c). A 532 nm laser is coupled into the central fiber through a 99:1 fiber splitter. Fluorescence returning through the outer fibers is passed through a 650 nm long-pass filter and detected with an auto-balanced photodetector, while the tap arm of the splitter provides the detector reference for laser-noise cancellation. Microwave signals are synthesized digitally on a cost-efficient Red Pitaya FPGA platform at an intermediate frequency, upconverted with an IQ mixer and local oscillator, amplified and delivered to the sensor head. This architecture keeps most optical and electronic components outside the measurement region while the sensor tip remains mechanically simple and compact. Apart from the initial laser-coupling path, the optical system remains completely fiber-based, which improves modularity and portability of the device.

\subsection*{Real-time magnetic field tracking}

Magnetic fields are measured through continuous-wave ODMR. Changes in the magnetic-field projection along an addressed NV axis shift the microwave resonance frequency of that spin subensemble, which is read out through monitoring the spin-dependent fluorescence. For magnetic imaging, repeatedly acquiring full ODMR spectra at each position of the scanned image would be inefficient. We therefore operate the sensor with digital closed-loop frequency tracking, building on prior high-dynamic-range NV magnetometry schemes \cite{clevenson_robust_2018,wang_realization_2023}.

The FPGA generates a sinusoidally frequency-modulated microwave signal and demodulates the digitized photodetector signal with a phase-coherent digital reference. For demodulation, the signal is filtered by a cascaded-integrator-comb (CIC) decimator followed by a minimum-phase finite-impulse-response filter, giving a 2 kHz measurement bandwidth and an output sampling rate of 30.5 kHz (Supplementary Note 4). Around the central ODMR zero crossing, the in-phase lock-in signal is proportional to the detuning between the microwave carrier and the resonance frequency. A discrete integral controller updates the microwave carrier frequency so that this error signal remains close to zero, allowing continuous tracking of the magnetic-field signal.

To increase the ODMR contrast of the $^{14}$N NV ensemble, the FPGA generates a three-tone frequency comb separated by 2.158 MHz, so that all three hyperfine transitions are driven simultaneously \cite{barry_optical_2016,schloss_simultaneous_2018}. The same digital backend controls ODMR sweeps, lock-in detection and frequency tracking. This reduces the amount of electrical equipment required and provides deterministic timing between microwave generation, demodulation and feedback.

Compared with open-loop scanning measurements operated around a fixed microwave frequency \cite{zhou_imaging_2021,du_scanning_2025}, the closed-loop implementation has two advantages. First, it allows for larger dynamic range of magnetic-field variations during scanning, because the feedback keeps the microwave carrier centered on the resonance. Second, in open-loop measurements the discriminator slope is measured once at the ODMR zero crossing and then used to convert subsequent voltage changes into frequency shifts. This conversion becomes inaccurate if the ODMR contrast or linewidth changes locally or over time, because both quantities affect the slope of the lock-in signal. Closed-loop tracking does not suffer from this issue. This makes the magnetic-field measurements more robust for samples whose material composition or magnetic environment is not spatially homogeneous or stable over time, as is often the case in application-relevant measurements on packaged or assembled devices. In our measurements, the battery tabs represent such a perturbation. From the demodulated lock-in signal we observe linewidth broadening near the ferromagnetic nickel anode tab, while the non-magnetic aluminium cathode tab produces no comparable degradation. We attribute this broadening to local magnetic-field inhomogeneity caused by the nickel tab.

\subsection*{Magnetic-field sensitivity}

\begin{figure*}[!t]
    \centering
    \includegraphics[width=\linewidth]{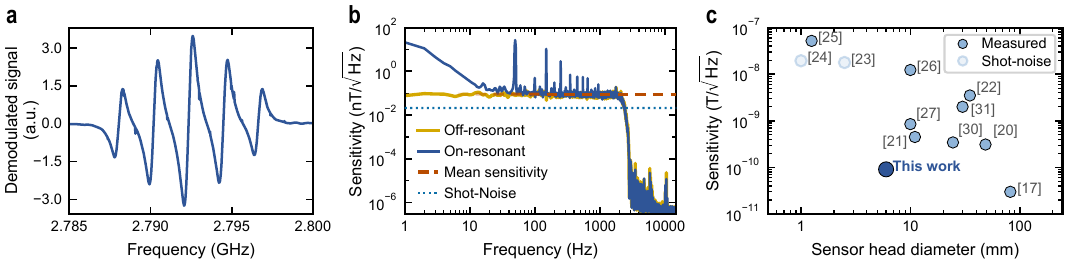}
    \caption{\textbf{Sensitivity characterization of the endoscopic NV magnetometer.} \textbf{a}, Lock-in ODMR spectrum for the $[100]$ bias-field alignment of the addressed $m_s=0 \leftrightarrow m_s=-1$ transition using simultaneous excitation of the $^{14}$N hyperfine triplet. The central zero crossing is used as the operating point for magnetic-field and sensitivity measurements. \textbf{b}, Magnetic-field amplitude spectral density for the $[100]$ bias-field alignment (blue). The brown dashed line indicates the mean sensitivity of $91 \, \mathrm{pT}/\sqrt{\mathrm{Hz}}$ between 10 Hz and 2 kHz after excluding known 50 Hz harmonics, and the dotted blue trace shows the measured shot-noise reference. The off-resonant trace is shown in yellow. \textbf{c}, Comparison of other reported fiber-coupled, magnetically unshielded NV magnetometers based on CW-ODMR in terms of magnetic-field sensitivity and sensor head diameter. Dark markers show measured sensitivities specified from the amplitude spectral density baseline, and light markers show studies for which only the shot-noise limit was specified.}
    \label{fig:sensitivity}
\end{figure*}

We first characterized the ODMR response and magnetic-field sensitivity of the sensor. The external bias magnetic field is aligned along the diamond $[100]$ direction, so that it projects equally onto all four NV orientations. This allows the $m_s=0 \leftrightarrow m_s=-1$ transitions, or equivalently the $m_s=0 \leftrightarrow m_s=+1$ transitions, of all four orientations to be addressed simultaneously. The fourfold increase in participating NVs improves the optical contrast by up to a factor four compared to a $[111]$ alignment. The reduced magnetic-field projection on each NV axis compared to the $[111]$ case is corrected by the geometric factor \(\sqrt{1/3}\) \cite{graham_fiber_coupled_2023, barry_optical_2016,alsid_solid-state_2023}. A representative lock-in ODMR spectrum shows the multi-feature response for simultaneous hyperfine excitation (Fig.~\ref{fig:sensitivity}a).

The open-loop sensitivity was determined from time traces acquired at the central ODMR zero crossing and converted to magnetic-field units using the measured frequency discriminator slope. We calculated double-sided magnetic-field power spectral densities from 16 independent 1 s traces, averaged the spectra in power and then took the square root to obtain the amplitude spectral density \cite{barry_sensitive_2024}. The reported sensitivity is the mean noise floor from 10 Hz to 2 kHz, excluding known 50 Hz mains harmonics. After optimizing the microwave power and modulation deviation over a two-dimensional parameter sweep, the sensor reaches a sensitivity of $91 \, \mathrm{pT}/\sqrt{\mathrm{Hz}}$ over this bandwidth (Fig.~\ref{fig:sensitivity}b and Supplementary Note 5). A corresponding measurement with the bias field aligned to a single $[111]$ NV orientation gives $169 \, \mathrm{pT}/\sqrt{\mathrm{Hz}}$ (Supplementary Note 5).

The measured sensitivity (see also Fig.~\ref{fig:sensitivity}b) remains above the shot-noise-limited sensitivity of $21.1 \, \mathrm{pT}/\sqrt{\mathrm{Hz}}$ (Supplementary Note 3), which identifies potential for future optimization of the sensor architecture. Further noise reduction may be achieved by suppressing technical noise contributions, such as detector noise and residual laser noise. Low-frequency noise could be reduced by improving vibration and temperature stability in the setup. The collected fluorescence remains approximately linear over the measured laser-power range, indicating that optical saturation of the excitation volume is not a limitation in the present operating regime (Supplementary Note 6). Together with the comparison in Fig.~\ref{fig:sensitivity}c, our measurements show that the sensor occupies a distinct region of the sensitivity--diameter landscape: it combines sub-100 pT/$\sqrt{\mathrm{Hz}}$ sensitivity with a 6 mm endoscopic sensor head and, to our knowledge, reaches the lowest reported noise floor for an endoscopic fiber-coupled NV sensor head below 10 mm diameter.

\subsection*{Battery magnetic-field imaging}

We used the endoscopic magnetometer to image the stray field of a commercial lithium-ion pouch cell during charging and discharging, which allows current distributions inside the cell to be monitored without electrical contact \cite{hu_sensitive_2020,bason_non-invasive_2022,brauchle_defect_2023,lee_diagnosis_2023,suzuki_non-destructive_2021,evans_quantum_2025,pollok_magnetic_2025}. The battery serves here as a representative demonstration of sensitive magnetic imaging on a packaged object in a crowded, unshielded setup. The cell, bias magnets, sensor, sample holder and translation stages must all be arranged within a limited volume, while the magnetic sensor has to remain close to the packaged cell. The endoscopic geometry allows the sensor head to be positioned in this measurement region, whereas the laser, photodetector and microwave-control electronics remain outside it. The scan area is currently limited primarily by the translation range of the positioning hardware, and the high sampling rate allows magnetic fields to be recorded continuously while the sample is moved.

\begin{figure}[!htbp]
    \centering
    \includegraphics[width=0.7\linewidth]{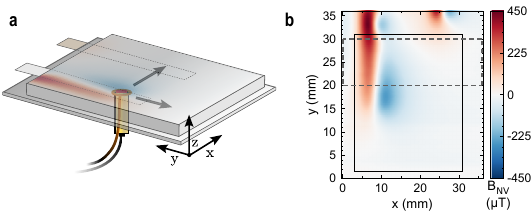}
    \caption{\textbf{Magnetic scanning of a lithium-ion pouch cell.} \textbf{a}, Measurement geometry. The pouch cell is translated across the fixed endoscopic sensor head with a motorized stage while the sensor measures the magnetic field. \textbf{b}, Zero-current map of the field $B_{\mathrm{NV}}$ with the battery resting at 3.7 V. The black outline marks the cell body, and the grey dashed box indicates the region scanned during charge and discharge measurements. The strong vertical feature on the left arises from the ferromagnetic nickel anode tab.}
    \label{fig:battery_setup}
\end{figure}

The pouch cell was mounted on a motorized $xy$ stage and scanned across the fixed sensor head at a stand-off distance of approximately 2 mm (Fig.~\ref{fig:battery_setup}a). The sensor measured the magnetic-field projection \(B_{\mathrm{NV}}=\hat{\mathbf{u}}\cdot\mathbf{B}\) along a single $[111]$ NV axis lying in the $xz$ plane of the battery coordinate system. A zero-current scan with the battery resting at 3.7 V reveals static magnetic features dominated by the ferromagnetic nickel anode tab and by magnetic signatures near the tab connectors (Fig.~\ref{fig:battery_setup}b). This map provides a reference for the cell geometry in the subsequent charge and discharge measurements.

During charge and discharge, we scanned a central strip of the cell while the FPGA tracking loop continuously followed the ODMR resonance. Each scan was corrected by subtracting a zero-current reference scan, which suppresses the static magnetic background from the tab materials and connectors. Four charging scans and four discharging scans were processed in this way and averaged. The resulting current-induced fields have opposite signs for charge and discharge, as expected from reversal of the current direction (Fig.~\ref{fig:current_reconstruction}, first column).

\subsection*{Current-density reconstruction}

The measured field maps were used to reconstruct the depth-integrated sheet-current density inside the battery. Because the stand-off distance is much larger than the separation between individual electrode layers, the measurement is insensitive to the exact depth of a given current path. We therefore model the cell as an effective current sheet and reconstruct \(\mathbf{K}(x,y)=\int\mathbf{J}(x,y,z)\,\mathrm{d}z\), with units of A/m \cite{bason_non-invasive_2022}. Previous magnetic current-imaging approaches have often used Fourier-based deconvolution or related regularized inverse methods \cite{roth_using_1989,feldmann_resolution_2004,broadway_improved_2020}. These methods are well suited when the measured field map covers the full source region. However, this is not always possible, such as when measurement duration is limited or the full source region is not spatially accessible. We therefore use a real-space regularized inversion \cite{brauchle_defect_2023, hofer_analyzing_2012, kishimoto_estimation_2016}. This approach also allows us to incorporate known physical priors directly on the spatial grid, such as prior knowledge about the spatial extent of the current distribution or current conservation. The full formulation, parameter choices and validation of our method are described in Supplementary Note 8.

\begin{figure}[!htbp]
    \centering
    \includegraphics[width=\linewidth]{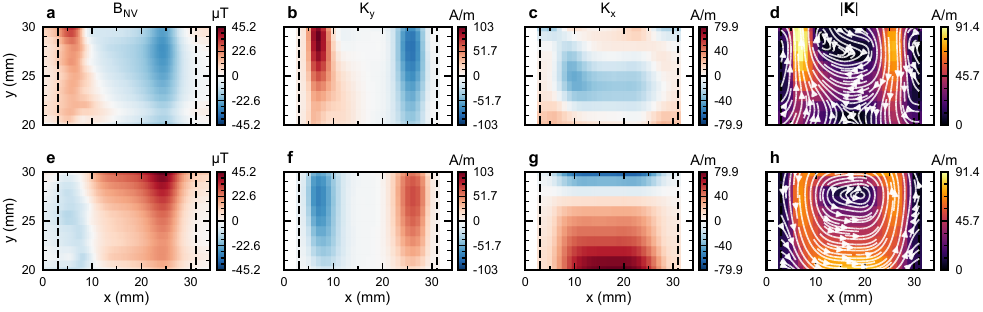}
    \caption{\textbf{Magnetic-field maps and reconstructed sheet-current density during battery operation.} Panels \textbf{a--d} show the charging measurements and panels \textbf{e--h} show the discharging measurements. The first column shows the measured background-subtracted magnetic field. The second and third columns show the reconstructed sheet-current components $K_y$ and $K_x$. The fourth column shows the current-density magnitude $|\mathbf{K}|$, with streamlines indicating the large-scale current flow. The reconstruction uses the real-space regularized inversion described in Supplementary Note 8.}
    \label{fig:current_reconstruction}
\end{figure}

The reconstructed maps show large-scale current flow consistent with the battery geometry (Fig.~\ref{fig:current_reconstruction}). During charging, current spreads from the cathode-tab side and converges toward the anode-tab side. During discharge, the pattern reverses. The strongest current densities occur near the tab region, while the central cell area is dominated by current paths connecting the tabs. Maps of the residual field after reconstruction and line profiles of the fields across the battery are shown in Supplementary Fig. 10. The reconstruction provides a macroscopic, depth-integrated current map at the few-millimeter length scale set by the measurement stand-off. This enables identification of large-scale current pathways inside the packaged cell.

Together, our results establish an endoscopic diamond magnetometer that combines a 6 mm diameter fiber-coupled sensor head, a remote optical and electronic backend, and FPGA-based real-time frequency tracking. The magnetometer achieves a sensitivity of $91 \, \mathrm{pT}/\sqrt{\mathrm{Hz}}$ over a 2 kHz bandwidth in an unshielded environment, while retaining a probe geometry suitable for sensing in spatially crowded and confined settings. Applied to a commercial lithium-ion pouch cell, the sensor enables non-contact, non-destructive measurement of macroscopic internal current-flow patterns through magnetic-field mapping during charge and discharge. This capability arises from the combination of miniaturization, sensitivity and deployability in a single platform: the fused fiber bundle and micro-objective allow excitation and fluorescence collection to be optimized separately inside a millimeter-scale sensor head, while the optical and electronic backend remains outside the measurement region. Our device also offers routes for further development. The measured shot-noise reference shows that improved suppression of technical noise sources can further increase sensitivity within the same platform. Pulsed ODMR protocols could also enhance sensitivity and bandwidth compared to the CW method used here \cite{barry_sensitive_2024}. Simultaneous tracking of multiple ODMR resonances would enable vector magnetic-field reconstruction together with temperature measurements during the same scan \cite{schloss_simultaneous_2018,shim_multiplexed_2022}. Further optimization could also reduce the stand-off distance between sample and sensor, which was limited mainly by broadening of the ODMR resonance near the ferromagnetic nickel tab as the sensor approached the cell. This would allow higher spatial resolution and thus reconstruction of finer current features inside the cell. More broadly, the combination of high sensitivity, compact probe geometry and real-time tracking makes our platform well suited for magnetic-field imaging in a variety of application-relevant settings beyond battery measurements, such as non-destructive testing of electronic components or magnetic imaging in biomedical contexts.

\section*{Methods}

\subsection*{Sensor head and optical setup}

The sensor head contains an isotopically purified diamond sample (>99.99\% $^{12}$C) with $[NV] = 4.5 \, \mathrm{ppm}$ and a $T_2^* \approx 1 \, \mathrm{\mu s}$ (DNV-B14, Element Six), a fused concentric fiber bundle (Castor Optics) and a custom-built micro-objective. The central excitation fiber has a 14 \textmu m core diameter and a numerical aperture of 0.12. Four surrounding fibers, each with a 200 \textmu m core diameter and a numerical aperture of 0.22, are packaged into the same bundle and collect fluorescence. The excitation fiber is spliced to a single-mode fiber for laser coupling. The micro-objective consists of a 0.25-pitch gradient-index lens (GRINTECH) and a 1 mm diameter, 0.83-NA spherical lens (Edmund Optics 35-710). The optical components are packaged into a ferrule machined from polyether ether ketone (PEEK), with the diamond positioned at the distal end of the micro-objective. Microwave fields are applied to the diamond by a loop antenna formed from a wire wound around the distal end of the sensor head, adjacent to the diamond. The diameter of the sensor head at its widest point is 6 mm.

Green excitation light from a 532 nm laser (Coherent Verdi V6) is coupled into the excitation fiber through a custom fiber assembly (Castor Optics) that combines a 1\(\times\)2 99:1 single-mode fiber splitter with the 4\(\times\)1 concentric fiber bundle. Typical operation used 800 mW of laser power before the fiber-coupling path, with approximately 60--65\% coupled into the input fiber. Collected fluorescence from the outer fibers passes through an in-line 650 nm long-pass filter and is detected with an auto-balanced photodetector (Newport Nirvana 2007). The tap arm of the splitter samples a portion of the laser light, which is sent to the detector reference input, providing common-mode suppression of laser-intensity noise. We employ a variable fiber attenuator between the fiber-splitter tap arm and the detector to optimize the power ratio between signal and reference for maximum common-mode rejection in the auto-balanced detection scheme. The measured fluorescence power at 800 mW, used for the shot-noise-limited sensitivity calculation, was 350 \textmu W (Supplementary Note 3).

\subsection*{Microwave generation and ODMR operation}

Microwave control tones are generated on a Red Pitaya STEMlab 125-14-LN FPGA platform by direct digital synthesis at an intermediate frequency. The signal is upconverted with an IQ mixer (Marki Microwave MMIQ-0218LXPC) and a local oscillator (Windfreak Technologies SynthNV Pro), and then amplified (Mini-Circuits ZHL-16W-43-S+) before delivery to the sensor head. For lock-in ODMR, the microwave frequency is sinusoidally modulated, with an instantaneous frequency \(f(t)=f_0+f_{\mathrm{dev}}\sin(2\pi f_{\mathrm{mod}}t)\), where \(f_0\) is the carrier frequency, \(f_{\mathrm{dev}}\) is the frequency deviation and \(f_{\mathrm{mod}}\) is the modulation frequency. The modulation frequency was set to 15.26 kHz, matched to the first null of the CIC filter, so that residual modulation-frequency components at the harmonics of \(f_{\mathrm{mod}}\) are already suppressed in the first decimation stage. For the $[100]$ sensitivity spectrum shown in Fig.~\ref{fig:sensitivity}b, the modulation deviation was $f_{\mathrm{dev}}=250 \, \mathrm{kHz}$ and the nominal Red Pitaya IF output-power setting was $-6.50 \, \mathrm{dBm}$ per hyperfine tone, corresponding to $0.224 \, \mathrm{mW}$ per tone into $50 \, \Omega$ before upconversion and amplification.

For $^{14}$N NV centers, three tones separated by 2.158 MHz are generated to drive the hyperfine triplet simultaneously, increasing the ODMR contrast and lock-in slope compared with single-hyperfine driving \cite{ahmadi_pump-enhanced_2017}. The bias field was generated by external ring magnets and aligned either along the diamond $[100]$ direction or along a single $[111]$ NV axis. For the final magnet geometry, simulations give a peak-to-peak variation of \(|\mathbf{B}|\) across the $3 \, \mathrm{mm}\times3 \, \mathrm{mm}\times0.6 \, \mathrm{mm}$ diamond of \(9.4 \, \mathrm{nT}\) or \(1.7\) ppm relative to the mean bias field. The $[100]$ alignment was used for the highest-sensitivity characterization, while the battery scans used the projected measurement axis \(\hat{\mathbf{u}}=(\sqrt{2/3},0,\sqrt{1/3})\) in the battery coordinate system. The bias-field homogeneity and magnet arrangement are described in Supplementary Note 7.

\subsection*{Digital lock-in detection and tracking}

The FPGA software builds on the PyRPL project as a general communication framework between the FPGA and PC \cite{neuhaus_python_2024}, while the microwave synthesis, digital lock-in demodulation and frequency-tracking modules are custom implementations developed for this work. The photodetector signal is digitized by the 14-bit analog-to-digital converter on the FPGA and multiplied with a phase-coherent digital reference synchronized to the microwave modulation. The signal is then processed by a two-stage filter chain. The first stage is a five-stage CIC decimator with decimation factor 4096 and differential delay 2, reducing the sampling rate from 125 MHz to 30.5 kHz. The second stage is a minimum-phase finite-impulse-response filter designed to compensate the passband droop of the first stage and define the 2 kHz measurement bandwidth. Filter responses and implementation details are given in Supplementary Note 4.

The demodulated in-phase signal is used as the error signal for closed-loop tracking, following the frequency-locking approach of Clevenson et al. \cite{clevenson_robust_2018}. Near the resonance, this signal is approximately linear in the frequency detuning, \(e[n]\approx K(f_0[n]-f_r[n])\), where $K$ is the discriminator slope, $f_0$ is the microwave carrier frequency and $f_r$ is the resonance frequency. The controller updates the direct-digital-synthesis tuning word according to \(f_0[n+1]=f_0[n]-\mu e[n]\), where $\mu$ is the integral gain. This feedback keeps the microwave frequency locked to the ODMR resonance during dynamic measurements.

\subsection*{Sensitivity measurement}

The magnetic-field sensitivity is defined as \(\eta=\delta B\sqrt{T}\), where \(\delta B\) is the smallest resolvable magnetic-field change after a measurement time \(T\) \cite{taylor_high_sensitivity_2008,bal_ultrasensitive_2012}. Typically the sensitivity is specified as a function of frequency \(\eta(f)\), i.e., as the amplitude spectral density (ASD) of the magnetic-field noise. A representative value for the sensitivity is then given as an average over frequency bands of interest, excluding low-frequency noise and known environmental signals, such as \(50 \, \mathrm{Hz}\) harmonics from mains-related interference \cite{schloss_simultaneous_2018,graham_fiber_coupled_2023,barry_sensitive_2024}. The sensitivity was measured in open-loop operation. First, a lock-in ODMR spectrum was recorded, and a local linear fit around the central zero crossing was used to determine the discriminator slope \(S_{\nu}\) between ADC units and microwave frequency. Time traces \(v(t)\) acquired at this operating point were converted to magnetic-field units using \(B(t)=v(t)/(S_{\nu}\gamma_e p)\), where \(p\) is the projection factor between the measured magnetic-field direction and the addressed NV axis. For the $[100]$ bias-field alignment \(p=1/\sqrt{3}\), and \(p=1\) for the single-orientation $[111]$ alignment \cite{graham_fiber_coupled_2023,barry_optical_2016}. For each sensitivity measurement, 16 independent 1 s traces were processed into double-sided power spectral densities and averaged in power before taking the square root to obtain the amplitude spectral density, following the convention that the reported sensitivity corresponds to the rms magnetic-field noise in a 1 Hz bin \cite{barry_sensitive_2024}. The reported sensitivity is the mean value between 10 Hz and 2 kHz after masking all 50 Hz harmonics. The resulting values are consistent with the time-domain definition \(\eta=\sigma_B/\sqrt{2f_{\mathrm{ENBW}}}\), where \(\sigma_B\) is the rms standard deviation of equivalently filtered magnetic-field time traces and \(f_{\mathrm{ENBW}}\) is the equivalent noise bandwidth \cite{schloss_simultaneous_2018}. The off-resonant trace in Fig.~\ref{fig:sensitivity}b was acquired under otherwise identical conditions but with the microwave drive detuned by around 20 MHz from resonance.

\subsection*{Battery scanning}

The battery measurements used a commercial lithium-ion pouch cell (MyLipo Q25-0550-E). The cell contains 19 graphite anode layers and 18 lithium cobalt oxide cathode layers separated by polymer films and enclosed in an aluminum-laminate pouch. The anode and cathode tabs were nickel and aluminum, respectively. The cell was mounted on a 3D-printed holder and translated across the fixed sensor head with two motorized linear stages (Thorlabs MTS50/M-Z8) in an $xy$ configuration. The stand-off distance between sensor head and cell surface was approximately 2 mm. A zero-current scan of the full cell was acquired at 3.7 V to identify static magnetic background features. Dynamic measurements were restricted to a strip covering \(x\in[0,36]\) mm and \(y\in[20,30]\) mm to limit acquisition time. During these measurements, the charging current remained close to \(560 \, \mathrm{mA}\), while the discharge current decreased from approximately \(570 \, \mathrm{mA}\) to \(550 \, \mathrm{mA}\).

During a scan, the sensor continuously sampled the magnetic field while the stage moved line by line along $x$ at \(2 \, \mathrm{mm/s}\) and stepped in $y$. The output sampling rate after lock-in decimation was 30.5 kHz, and the data were binned into a $100\times15$ pixel map according to stage position. Charging and discharging maps were each corrected by subtracting a zero-current reference scan, and four scans were averaged for each direction. Mild Gaussian smoothing with \(\sigma=(1.5,1)\) pixels was applied before current reconstruction to suppress scan-to-scan noise from small stage offsets.

\subsection*{Current-density reconstruction}

The battery was modeled as an effective current sheet because the 2 mm stand-off is significantly larger than the electrode-layer spacing. The forward model relates the sheet current \(\mathbf{K}=(K_x,K_y)\) to the measured projected field \(B_{\mathrm{NV}}\) through the Biot-Savart law. The inverse problem was solved on a 1 mm source grid with a regularized real-space least-squares method. The source grid covered \(x\in[3,31]\) mm and \(y\in[17,33]\) mm; the measured strip was padded by 3 mm in $y$ so that currents outside the directly measured window could still contribute to fields near the scan boundary in the reconstruction model. The objective function included a data-fidelity term, a smoothness penalty, a divergence penalty, a row-wise flux-balance penalty and a boundary tapering penalty. The smoothness parameter was selected separately for charge and discharge by an L-curve criterion, while the other penalty weights were trace-normalized and fixed. The complete formulation, parameter choices and residual field maps are provided in Supplementary Note 8.

\section*{Data availability}

The data supporting the findings of this study are available from the corresponding authors upon reasonable request.

\section*{Code availability}

The code supporting the findings of this study is available from the corresponding author upon reasonable request.

\section*{Acknowledgements}

We thank Andre Pointner, Tim Ascherl and Gavin Morley for fruitful discussions.

\section*{Funding}

J.W. and R.N. disclose support for the research of this work from the European Union under the Key Digital Technologies Joint Undertaking (KDT JU), now Chips Joint Undertaking (Chips JU), in the projects ARCHIMEDES (grant nos. 16MEE0329 and 101112295) and MOSAIC (grant nos. 16MEE0494 and 101081238), and from the German Federal Ministry of Education and Research (BMBF, now BMFTR) in the projects INNOBAT (grant no. 03XP0492D) and QuaLiProM (grant no. 03XP0573C).

\section*{Author contributions}

J.W. designed and built the sensor system, performed the measurements, analyzed the data and prepared the figures. R.N. supervised the project, contributed to the experimental design and evaluation, and acquired funding. Both authors discussed the results, contributed to interpretation of the data and wrote the manuscript.

\section*{Competing interests}

J.W. and R.N. are inventors on a patent application related to the optical excitation and collection scheme described in this work (application number DE 10 2024 132 240.3).

\clearpage
\newgeometry{left=0.85in,right=0.85in,top=0.9in,bottom=0.9in}
\setcounter{figure}{0}
\setcounter{equation}{0}
\renewcommand{\theequation}{S\arabic{equation}}
\renewcommand{\thefigure}{S\arabic{figure}}
\renewcommand{\theHequation}{S\arabic{equation}}
\renewcommand{\theHfigure}{S\arabic{figure}}
\renewcommand{\figurename}{Supplementary Fig.}
\renewcommand{\refname}{Supplementary References}
\setlength{\parskip}{0.45em}
\setstretch{1.03}
\titleformat{\section}{\large\bfseries}{\thesection}{0.75em}{}
\titlespacing*{\section}{0pt}{1.5ex plus 0.4ex minus 0.2ex}{0.8ex}
\titleformat{\subsection}{\normalsize\bfseries}{\thesubsection}{0.75em}{}
\titlespacing*{\subsection}{0pt}{1.1ex plus 0.3ex minus 0.2ex}{0.4ex}
\renewcommand{\contentsname}{Contents}
\renewcommand{\cfttoctitlefont}{\large\bfseries}
\renewcommand{\cftaftertoctitle}{\par\vspace{0.3em}}
\renewcommand{\cftsecfont}{\normalsize}
\renewcommand{\cftsecpagefont}{\normalsize}
\setlength{\cftbeforesecskip}{0.2em}
\setlength{\cftsecindent}{0pt}
\setlength{\cftsecnumwidth}{0pt}

\begin{center}
{\large\bfseries Supplementary Information\par}
\vspace{0.5em}
{\normalsize Sensitive endoscopic diamond magnetometer for non-contact sensing in confined environments \par}
\vspace{1em}
Johannes Wesseler$^{1}$ and Roland Nagy$^{1,*}$\par
\vspace{0.5em}
$^{1}$Institute of Applied Quantum Technologies, Friedrich-Alexander-Universit\"at Erlangen-N\"urnberg, Erlangen, Germany\par
$^*$Correspondence: roland.nagy@fau.de
\end{center}

\tableofcontents
\clearpage

\section*{Supplementary Note 1: NV-center spin physics and ODMR magnetometry}
\addcontentsline{toc}{section}{Supplementary Note 1: NV-center spin physics and ODMR magnetometry}

The negatively charged nitrogen-vacancy center (NV$^{-}$) in diamond is a point defect formed by a substitutional nitrogen atom next to a vacancy in the diamond lattice \cite{si_doherty_theory_2012}. The defect axis lies along one of the four crystallographic \(\langle 111\rangle\) directions. The electronic ground state is a spin triplet with a zero-field splitting of \(D\approx 2.87 \, \mathrm{GHz}\) between the \(m_s=0\) and \(m_s=\pm 1\) sublevels \cite{si_doherty_nitrogen_vacancy_2013}. Optical excitation drives the system predominantly spin-conserving to an excited triplet state with a zero-phonon line at 637 nm, however the excitation is typically performed off-resonantly with a green laser at 532 nm. Fluorescence is emitted during radiative decay from the excited state in the red and near-infrared, with a broad phonon sideband extending to approximately 800 nm. A metastable singlet manifold provides an additional non-radiative decay pathway back to the ground state. Because this intersystem crossing is more likely for the \(m_s=\pm 1\) states than for \(m_s=0\), continuous green illumination can polarize the spin preferentially into \(m_s=0\) and produces spin-dependent red fluorescence. Microwave fields resonant with the spin transitions then transfer population out of \(m_s=0\), reduce the fluorescence, and thereby enable optically detected magnetic resonance (ODMR), that is, measurement of the spin state and resonance frequencies through the fluorescence intensity. The spin resonance frequencies are sensitive to magnetic fields via the Zeeman effect, so ODMR can be used for magnetometry.

\begin{figure}[htbp]
    \centering
    \scalebox{0.75}{%
        \begin{minipage}{\linewidth}
            \centering
            \raisebox{-0.53\height}{\includegraphics[width=0.43\linewidth]{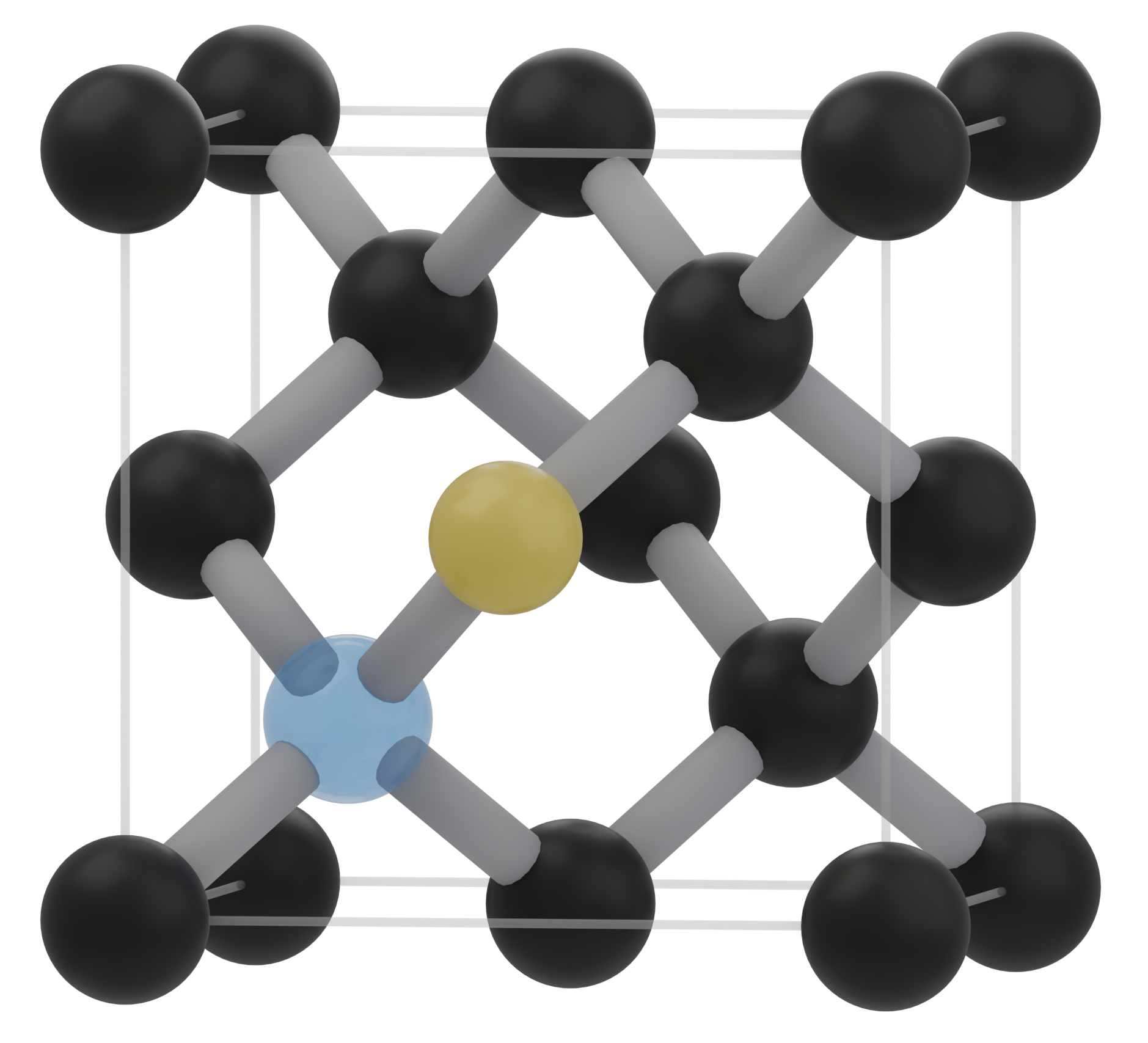}}%
            \hfill
            \raisebox{-0.5\height}{\includegraphics[width=0.48\linewidth]{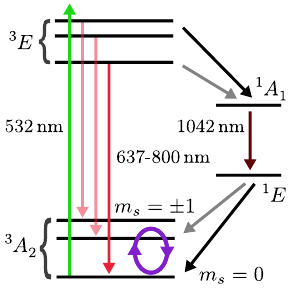}}%
        \end{minipage}%
    }
    \caption{\textbf{NV center and energy level diagram.} Left, schematic of the nitrogen-vacancy defect in the diamond lattice. Right, electronic level diagram showing optical excitation and fluorescence, microwave-driven spin transitions and the non-radiative decay path responsible for spin-dependent fluorescence.}
    \label{fig:si_nv}
\end{figure}

The NV$^{-}$ ground-state Hamiltonian for one NV orientation can be approximated as
\begin{equation}
\frac{H}{h}\approx D S_z^2+\mathbf{S}\cdot\mathbf{A}\cdot\mathbf{I}+\gamma_e \mathbf{B}\cdot\mathbf{S}
\end{equation}
where \(D\approx 2.87\) GHz is the zero-field splitting, \(\mathbf{A}\) is the hyperfine tensor coupling the electron spin to the nitrogen nuclear spin, and \(\gamma_e\approx 28\) MHz/mT is the electron gyromagnetic ratio in frequency units \cite{si_doherty_theory_2012,si_doherty_nitrogen_vacancy_2013}. \(\mathbf{A}=\operatorname{diag}(A_{\perp},A_{\perp},A_{\parallel})\), giving \(\mathbf{S}\cdot\mathbf{A}\cdot\mathbf{I}=A_{\perp}(S_xI_x+S_yI_y)+A_{\parallel}S_zI_z\). The transverse hyperfine term couples states with different electronic spin projection and is neglected in the secular approximation.

For NV orientation \(i\), the axial component of the magnetic field is the projection \(B_{\parallel,i}=\hat{\mathbf n}_i\cdot\mathbf B\) onto that NV axis. In the linear Zeeman regime, the allowed \(\Delta m_I=0\) ODMR transition frequencies are then approximately
\begin{equation}
\nu_{\pm,i,m_I}\approx D\pm \gamma_e B_{\parallel,i} +  m_IA_{\parallel},
\end{equation}
where the upper and lower signs refer to the \(m_s=0\leftrightarrow m_s=+1\) and \(m_s=0\leftrightarrow m_s=-1\) transitions, respectively, and \(m_I=0,\pm1\) for \(^{14}\mathrm N\). Adjacent hyperfine components are separated by \(|A_{\parallel}|\approx 2.158\) MHz. A small magnetic-field change therefore shifts the addressed transition by
\begin{equation}
\delta\nu_i \simeq \pm \gamma_e \hat{\mathbf n}_i\cdot\delta\mathbf B,
\end{equation}
with the sign determined by the addressed \(m_s=0\leftrightarrow m_s=+1\) or \(m_s=0\leftrightarrow m_s=-1\) transition. A bias field aligned along a single \([111]\) NV direction gives maximum projection for that orientation and separates its resonance from the other three orientations. A field orientation along \([100]\), in contrast, reduces the axial field component by \(1/\sqrt{3}\) for all four NV orientations. However, the \([100]\) geometry allows the spin transitions of all four orientation classes to be driven with a single microwave tone, increasing the contrast of that transition.

The magnetometer is operated in continuous-wave lock-in ODMR. The microwave carrier is sinusoidally frequency modulated around the resonance, and the fluorescence is demodulated at the modulation frequency. For modulation depths comparable to the ODMR linewidth, the in-phase lock-in signal has a dispersive line shape and is approximately proportional to the derivative of the ODMR fluorescence spectrum near resonance \cite{si_ahmadi_pump-enhanced_2017}. The central zero crossing is therefore a frequency discriminator: close to the operating point, a resonance-frequency shift caused by a magnetic field change produces a lock-in voltage proportional to that shift. The discriminator slope is measured experimentally from a lock-in ODMR sweep and is used to convert voltage signals into frequency changes and thus into magnetic-field signals.

Because the sample contains \(^{14}\)N NV centers with nuclear spin \(I=1\), each electronic transition consists of a three-line hyperfine triplet. Driving only one hyperfine component leaves two thirds of the population weakly addressed and reduces the available contrast. We therefore always synthesize three microwave tones separated by 2.158 MHz and modulate them together, so that the three hyperfine components are driven simultaneously. This increases the lock-in slope and improves the magnetic-field sensitivity without other changes to the measurement protocol \cite{si_barry_optical_2016}.

\clearpage
\section*{Supplementary Note 2: Optical simulations}
\addcontentsline{toc}{section}{Supplementary Note 2: Optical simulations}

We simulated the optical excitation and collection paths of the sensor head with Ansys Zemax OpticStudio in non-sequential ray-tracing mode. The model includes the central excitation fiber, the gradient-index lens, the spherical lens and the diamond. The excitation fiber has a 14 \textmu m core diameter and a numerical aperture of 0.12. The central fiber is imaged onto the diamond by the micro-objective, while fluorescence generated in the diamond is collected in the reciprocal direction by the four outer fibers of the bundle.

\begin{figure}[htbp]
    \centering
    \includegraphics[width=0.7\linewidth]{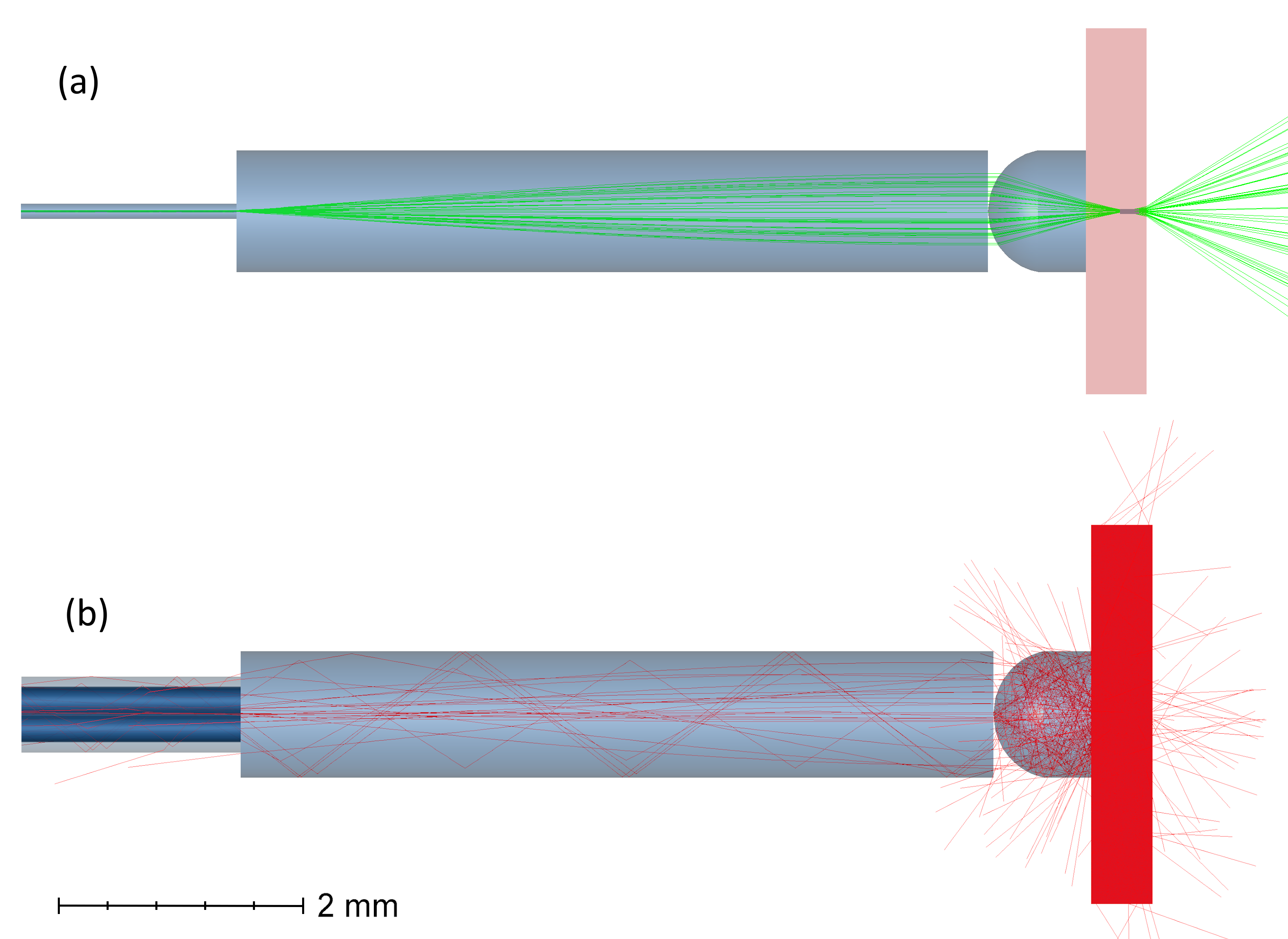}
    \caption{\textbf{Ray-tracing simulation of the sensor head optics.} The optical model includes the fiber bundle, micro-objective and diamond. \textbf{a} The excitation path focuses light into the diamond. \textbf{b} Emitted rays are traced from the excitation region to estimate the fraction collected by the outer fibers.}
    \label{fig:si_zemax_render}
\end{figure}

To quantify the extent of the excitation region, we extracted the simulated optical flux profile around the focus. Because the central fiber is strongly multimode at 532 nm, the intensity distribution is not well described by a Gaussian. We therefore characterize the lateral spot size by its full width at half maximum (FWHM) instead. The simulated lateral FWHM is 2.9 \textmu m. The axial extent is characterized by the depth of focus, defined as the axial range over which the lateral FWHM remains within a factor of $\sqrt{2}$ of its minimum value. This gives a depth of focus of 53 \textmu m. Approximating the excitation region as a cylinder gives a volume of approximately $350 \, \mu\mathrm{m}^3$.

\begin{figure}[htbp]
    \centering
    \includegraphics[width=0.72\linewidth]{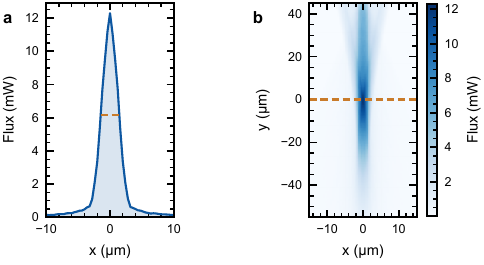}
    \caption{\textbf{Simulated excitation flux in the diamond.} \textbf{a} Lateral flux profile through the focal plane. The dashed line indicates the full width at half maximum of 2.9 \textmu m. \textbf{b} Two-dimensional flux map through the focus. The dashed line indicates the cross-section along that \textbf{a} is taken. Coordinates are centered on the peak intensity.}
    \label{fig:si_zemax_flux}
\end{figure}

For the collection-efficiency simulation, we placed an isotropic cylindrical volume source at the simulated focus. The source diameter and height were set to 2.9 \textmu m and 53 \textmu m, respectively. Rays were traced through the diamond and micro-objective, and rays impinging on the outer fibers at angles above their acceptance angle were excluded. This model gives an approximate fluorescence collection efficiency of 2.5\%. The estimate does not include possible imperfections in alignment or non-concentricity of the fiber-cores or surface-imperfections, but it captures the benefit of separating excitation and collection fibers and employing the high-NA micro-objective for increased collection efficiency.

\clearpage
\section*{Supplementary Note 3: Shot-noise limited sensitivity}
\addcontentsline{toc}{section}{Supplementary Note 3: Shot-noise limited sensitivity}

\begin{figure}[htbp]
    \centering
    \includegraphics[width=0.54\linewidth]{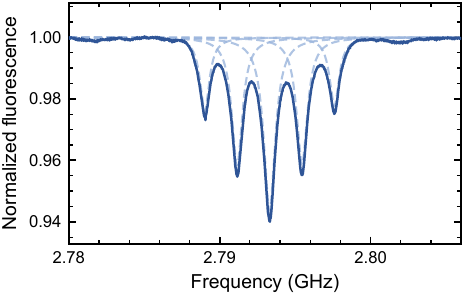}
    \caption{\textbf{ODMR spectrum for shot-noise estimation.} The dashed lines show Lorentzian fits used to extract the linewidth and contrast. The measurement was performed for the $[100]$ bias-field alignment.}
    \label{fig:si_cw_odmr}
\end{figure}

In the shot-noise limit of continuous-wave ODMR magnetometry, the magnetic-field sensitivity can be determined as
\begin{equation}
    \eta_{\mathrm{SN}}=\frac{4}{3\sqrt{3}}\frac{\Delta \nu}{\gamma_e C \sqrt{R}},
\end{equation}
where $\Delta \nu$ is the ODMR linewidth, $C$ is the ODMR contrast, $\gamma_e$ is the electron gyromagnetic ratio and $R$ is the detected photon rate \cite{si_dreau_avoiding_2011}. To estimate this limit for the sensor, we recorded a continuous-wave ODMR spectrum without lock-in demodulation and balanced detection. As for the sensitivity measurements in the main text, the bias magnet was aligned along a $[100]$ direction, such that the resonances of the four NV orientations overlap. The fluorescence was measured on an optical power meter (MKS Newport 843-R), with the detection wavelength set at 750 nm. With the microwaves tuned away from resonance, we measured $P_{\mathrm{opt}}=350 \, \mu\mathrm{W}$ at this setting. Using the manufacturer-calibrated responsivity curves for the power meter and the normalized photoluminescence spectra provided for another DNV-B14 sample in \cite{si_yang_light_2025}, we determined a photon rate of $1.34 \times 10^{15} \, \mathrm{s}^{-1}$. The ODMR spectrum with normalized fluorescence is shown in Figure \ref{fig:si_cw_odmr}. The ODMR spectrum was fitted with a sum of five offset Lorentzians to extract the linewidth and contrast. For the center resonance, the fitted linewidth was $\Delta\nu=0.91 \, \mathrm{MHz}$ and the contrast was $C=5.6\%$. With the measured photon rate and including again the angle factor of $1/\sqrt{3}$ due to the [100] alignment \cite{si_graham_fiber_coupled_2023}, these values correspond to a shot-noise-limited sensitivity of $21.1 \, \mathrm{pT}/\sqrt{\mathrm{Hz}}$.

\clearpage
\section*{Supplementary Note 4: FPGA lock-in filter chain}
\addcontentsline{toc}{section}{Supplementary Note 4: FPGA lock-in filter chain}

\begin{figure}[!htbp]
    \centering
    \makebox[\linewidth][c]{%
        \begin{minipage}[t]{0.62\linewidth}
            \vspace{0pt}
            \makebox[\linewidth][c]{%
                \includegraphics[width=0.98\linewidth]{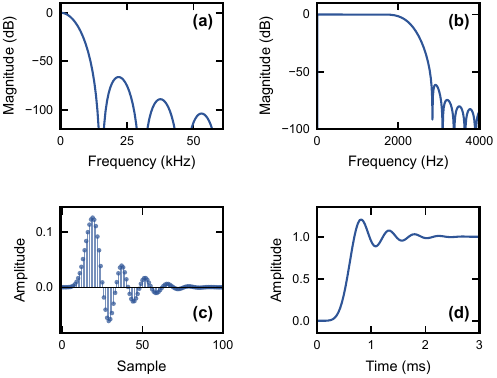}%
            }
        \end{minipage}%
        \hspace{0.075\linewidth}%
        \begin{minipage}[t]{0.24\linewidth}
            \vspace{0pt}
            \makebox[\linewidth][c]{%
                \includegraphics[width=0.98\linewidth]{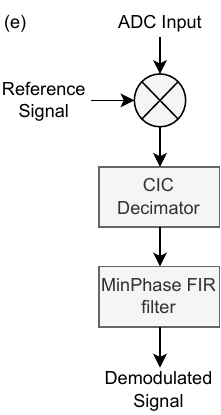}%
            }
        \end{minipage}%
    }
    \caption{\textbf{Digital filter responses and signal chain for FPGA lock-in detection.} \textbf{a}, Magnitude response of the first-stage CIC decimator. \textbf{b}, Magnitude response of the cascaded CIC and minimum-phase FIR filter. \textbf{c}, Impulse response of the cascaded filter chain at the decimated output rate. \textbf{d}, Corresponding step response. \textbf{e}, Schematic of the FPGA demodulation chain showing the signal multiplication and the CIC decimator followed by the minimum-phase FIR filter.}
    \label{fig:si_filters}
\end{figure}

The FPGA implements the lock-in demodulation, filtering and resonance-tracking feedback in real time. The signal from the photodetector is digitized by the 14-bit analog-to-digital converter at 125 MHz and multiplied with a 17-bit phase-coherent digital reference. The demodulated signal is then processed by a two-stage decimation and low-pass chain (see Figure \ref{fig:si_filters}). The first stage is a cascaded-integrator-comb (CIC) decimator with decimation factor $R=4096$, which reduces the sampling rate from $f_s=125 \, \mathrm{MHz}$ to
\(f_s'=f_s/R\approx 30.5 \, \mathrm{kHz}\).

For an $N$-stage CIC implementation with differential delay $M$, the magnitude response follows the sinc envelope
\begin{equation}
    |H_{\mathrm{CIC}}(f)| \propto
    \left|\frac{\sin(\pi f R M/f_s)}{\sin(\pi f/f_s)}\right|^N .
\end{equation}
The implemented decimator uses $N=5$ and $M=2$, so the first comb null occurs at $f_s/(RM)\approx 15.26 \, \mathrm{kHz}$, which is also the modulation frequency. The CIC stage therefore suppresses components at the modulation frequency and harmonics before the signal reaches the second filter.

The second stage is a minimum-phase finite-impulse-response filter at the reduced sampling rate. It compensates the CIC passband droop over the 2 kHz measurement bandwidth and sharpens the roll-off. The filter responses were designed in MATLAB and implemented on the FPGA using the Xilinx CIC and FIR compiler cores. The implemented FIR has 119 taps (order 118), and the coefficients are quantized to 18-bit signed values. The 18-bit quantization changes the FIR passband magnitude by less than \(4\times10^{-5}\,\mathrm{dB}\) compared with the floating-point design. The minimum-phase design front-loads the impulse response and reduces the low-frequency group delay relative to a comparable linear-phase filter. For the implemented filter chain, the low-frequency group delay is approximately 16 output samples, or \(0.53\,\mathrm{ms}\), with \(0.16\,\mathrm{ms}\) from the CIC stage and \(0.37\,\mathrm{ms}\) from the FIR stage. This low latency is important because the filtered lock-in signal is also used as the feedback error signal for resonance tracking.

\clearpage
\section*{Supplementary Note 5: Sensitivity optimization and bias-field alignment}
\addcontentsline{toc}{section}{Supplementary Note 5: Sensitivity optimization and bias-field alignment}

\subsection*{Sensitivity optimization across microwave parameters}

\begin{figure}[htbp]
    \centering
    \includegraphics[width=7.5cm]{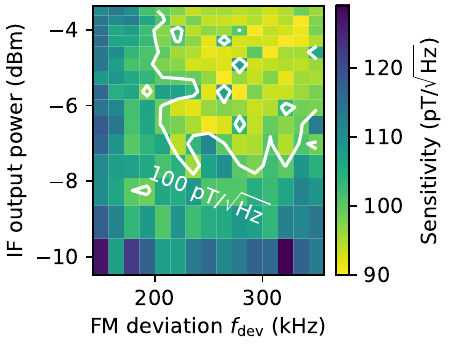}
    \caption{\textbf{Sensitivity optimization over microwave-power and frequency-modulation settings.} Magnetic-field sensitivity for the $[100]$ bias-field alignment measured as a function of frequency-modulation deviation $f_{\mathrm{dev}}$ and microwave power. The white contour marks the region below $100 \, \mathrm{pT}/\sqrt{\mathrm{Hz}}$.}
    \label{fig:si_sensitivity_optimization}
\end{figure}

To find the optimal working point for the magnetometer in terms of microwave power and FM deviation for the $[100]$ bias-field alignment, we measured the magnetic-field sensitivity over a two-dimensional sweep of those parameters (see Fig.~\ref{fig:si_sensitivity_optimization}). The modulation frequency was kept constant at 15.26 kHz, as we could not observe a significant sensitivity dependence on modulation frequency in earlier experiments and as this frequency sits at the first null of the CIC filter, see Supplementary Note 4. 

The sweep covered IF powers from $-10.0$ to $-3.5 \, \mathrm{dBm}$ on the Red Pitaya outputs and frequency deviations from $150$ to $350 \, \mathrm{kHz}$. The sensitivity generally improved with increasing microwave power and frequency deviation, but the optimum was broad, with large parts of the parameter space giving sensitivities below $100 \, \mathrm{pT}/\sqrt{\mathrm{Hz}}$. We did not increase the microwave power beyond $-3.5 \, \mathrm{dBm}$ to avoid excessive heating of the microwave antenna and due to significantly stronger spurs in the microwave spectrum at higher powers.

\subsection*{Sensitivity measurement for $[111]$ bias-field alignment}

\begin{figure}[htbp]
    \centering
    \includegraphics[width=0.5\linewidth]{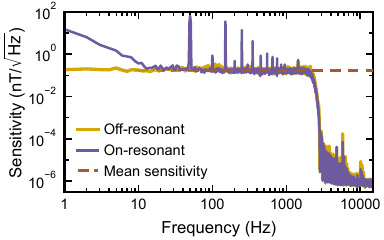}
    \caption{\textbf{Sensitivity for $[111]$ bias-field alignment.} The on-resonant amplitude spectral density is shown in purple. The brown dashed line indicates the mean sensitivity of $169 \, \mathrm{pT}/\sqrt{\mathrm{Hz}}$ between 10 Hz and 2 kHz after excluding 50 Hz harmonics. The off-resonant trace is shown in yellow.}
    \label{fig:si_sens111}
\end{figure}

In addition to the $[100]$ alignment used for sensitivity measurement in the main text, we characterized the sensor with the bias field aligned along a single $[111]$ NV axis. In this geometry, the bias field separates the ODMR spectrum into one transition pair associated with the NV orientation parallel to the field and a triply degenerate pair associated with the other three orientations. We probe the resonance of the aligned NV class, so only one quarter of the NV ensemble contributes to the ODMR contrast, but the magnetic-field projection onto the addressed NV axis is maximized. No geometric projection correction is therefore required. For the $[111]$ sensitivity measurement, the excitation polarization was also optimized to maximize the ODMR contrast of the measured resonance. The optimized sensitivity in this configuration was $169 \, \mathrm{pT}/\sqrt{\mathrm{Hz}}$ (see Fig.~\ref{fig:si_sens111}).

\clearpage
\section*{Supplementary Note 6: Fluorescence scaling with excitation power}
\addcontentsline{toc}{section}{Supplementary Note 6: Fluorescence scaling with excitation power}

\begin{figure}[htbp]
    \centering
    \includegraphics[width=0.6\linewidth]{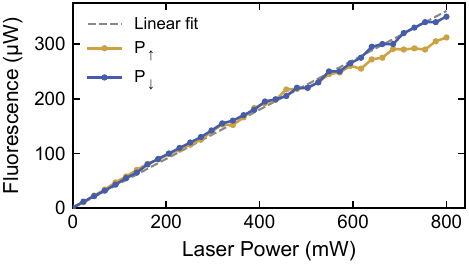}
    \caption{\textbf{Fluorescence-power scaling with excitation power.} Collected fluorescence power as a function of laser power before the fiber coupler for two measurement runs. \(P_{\uparrow}\) denotes the sweep from low to high laser powers with the fiber coupling aligned at low power, and \(P_{\downarrow}\) denotes the sweep from high to low powers with the coupling aligned at high power. The dashed line shows a linear fit through the origin to the \(P_{\downarrow}\) data.}
    \label{fig:si_saturation}
\end{figure}

The combination of the small-diameter, low-NA excitation fiber with the high-NA micro-objective focuses the excitation light into a small volume in the diamond, which could potentially lead to optical saturation. To evaluate this, we measured the collected fluorescence power as a function of the laser power before the fiber coupler (Fig.~\ref{fig:si_saturation}). The laser power was first swept from low to high values (\(P_{\uparrow}\)), with the fiber coupling aligned at low power, and then from high to low values (\(P_{\downarrow}\)), with the coupling aligned at high power. The \(P_{\downarrow}\) curve remains approximately linear over the full measured range of \(0\)--\(800\,\mathrm{mW}\). The divergence between the two sweeps above approximately \(500\,\mathrm{mW}\) is attributable to power-dependent drift in the fiber-coupling efficiency rather than optical saturation of the full excitation volume. While the full excitation volume does not show saturation under these conditions, individual NV centers near the center of the focus could still be locally saturated. At higher powers the variance between repeated fluorescence measurements also increases and the measured fluorescence fluctuates for tens of seconds after changing the laser power, which we attribute to thermal effects in the fiber coupling and sensor head. We conclude that optical saturation of the full excitation volume is not a limiting factor for the current sensitivity, but that thermal management and coupling stability will be important for further improvements at higher optical powers.

\clearpage
\section*{Supplementary Note 7: Bias magnetic field}
\addcontentsline{toc}{section}{Supplementary Note 7: Bias magnetic field}

This note quantifies the homogeneity of the ring-magnet bias field used for the ODMR measurements. The distance between the ring magnets was optimized in simulation for high field homogeneity across the diamond. The optimal distance differs from the Helmholtz condition because the magnets have finite dimensions rather than ideal current-loop geometry.

\begin{figure}[htbp]
    \centering
    \includegraphics[width=0.55\linewidth]{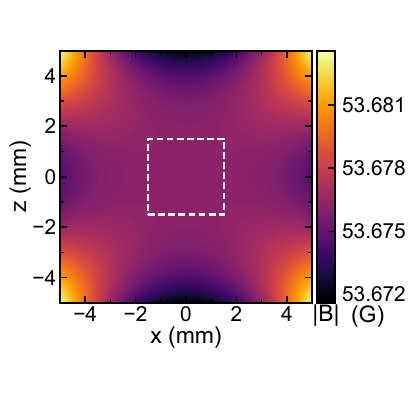}
    \caption{\textbf{Simulated bias magnetic field.} Magnetic-field magnitude along the axis between the ring magnets. Dashed lines indicate the extent of the diamond.}
    \label{fig:si_bias_field}
\end{figure}

For the 10 cm diameter rings used here, simulations with magpylib \cite{si_magpylib} give a mean field magnitude of \(53.68 \, \mathrm{G}\) and a peak-to-peak variation of \(9.4\times10^{-5} \, \mathrm{G}\) across the $3 \, \mathrm{mm} \times 3 \, \mathrm{mm} \times 0.6 \, \mathrm{mm}$ diamond for the chosen configuration (see Fig.~\ref{fig:si_bias_field}). This corresponds to \(94 \, \mu\mathrm{G}\), or \(1.7\) ppm relative to the mean bias field, and ensures that the ODMR linewidth is not dominated by macroscopic bias-field gradients across the sample.

\clearpage
\section*{Supplementary Note 8: Current-density reconstruction}
\addcontentsline{toc}{section}{Supplementary Note 8: Current-density reconstruction}

The battery scans reported in the main text are used to reconstruct the in-plane sheet-current density from the measured magnetic field. The sensor measures a magnetic stray-field map, while the quantity of interest is the underlying current density. The forward problem is well defined: if the current distribution is known, the magnetic field can be calculated from the Biot-Savart law. The inverse problem is ill-conditioned because each measured field value receives contributions from the entire current distribution and because the field is measured at finite stand-off \cite{si_roth_using_1989,si_feldmann_resolution_2004,si_broadway_improved_2020,si_midha_optimized_2024}. The finite stand-off also acts as a spatial low-pass filter: in Fourier space, a current-density mode with spatial wavenumber $k$ is attenuated at a measurement height $h$ approximately as $e^{-kh}$. Recovering the current therefore requires inverse amplification, so high-spatial-frequency noise is amplified together with any true fine structure.

The sensor was operated at a stand-off of approximately $h=2 \, \mathrm{mm}$, larger than the spacing between individual electrode layers. The measurement is therefore insensitive to the exact depth of a given current path, and we reconstruct the depth-integrated sheet current
\begin{equation}
    \mathbf{K}(x,y)=\int \mathbf{J}(x,y,z)\,\mathrm{d}z .
\end{equation}
Charge conservation implies $\partial_x K_x+\partial_y K_y=0$ because no current leaves the cell through its top or bottom surfaces. The measured magnetic-field projection is along
\begin{equation}
    \hat{\mathbf{u}}=\left(\sqrt{\frac{2}{3}},0,\sqrt{\frac{1}{3}}\right),
\end{equation}
in the coordinate system of the battery.

\begin{figure}[htbp]
    \centering
    \includegraphics[width=0.95\linewidth]{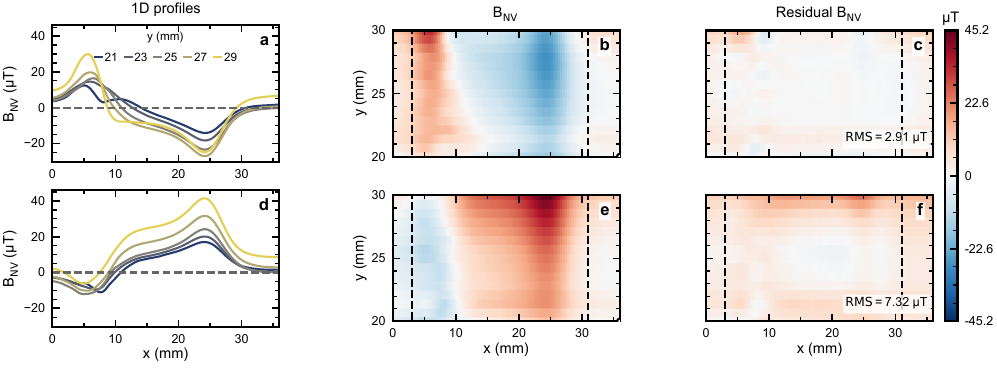}
    \caption{\textbf{Measured magnetic fields and reconstruction residuals.} \textbf{a,d}, One-dimensional profiles of the measured $B_{\mathrm{NV}}(x)$ field at selected $y$ positions for charge and discharge. \textbf{b,e}, Two-dimensional measured field maps. \textbf{c,f}, Residual field maps showing the difference between measured and reconstructed fields.}
    \label{fig:si_residuals}
\end{figure}

\subsection*{Discretized forward problem}

Following the sheet-current description used in magnetic current imaging of thin samples and battery cells \cite{si_roth_using_1989,si_brauchle_defect_2023}, the battery is modeled as an effective current sheet in the plane $z=0$. The Biot-Savart law for a surface current is
\begin{equation}
    \mathbf{B}(\mathbf{r})=\frac{\mu_0}{4\pi}
    \int \frac{\mathbf{K}(\mathbf{r}')\times(\mathbf{r}-\mathbf{r}')}{|\mathbf{r}-\mathbf{r}'|^3}\,\mathrm{d}A',
    \label{eq:si_biot_savart}
\end{equation}
where $\mathbf{r}=(x,y,h)$ is the observation point and $\mathbf{r}'=(x',y',0)$ is a source point. For a source pixel of area $\Delta A$ and offsets $\Delta x=x-x'$ and $\Delta y=y-y'$, with $r=\sqrt{\Delta x^2+\Delta y^2+h^2}$, the field contribution is
\begin{equation}
    \begin{aligned}
        \Delta B_x &= \frac{\mu_0}{4\pi}\frac{h K_y \Delta A}{r^3},\\
        \Delta B_y &= -\frac{\mu_0}{4\pi}\frac{h K_x \Delta A}{r^3},\\
        \Delta B_z &= \frac{\mu_0}{4\pi}\frac{(\Delta y K_x-\Delta x K_y)\Delta A}{r^3}.
    \end{aligned}
    \label{eq:si_pixel_components}
\end{equation}
The measured field is the projection
\begin{equation}
    B_{\mathrm{NV}}(x,y,h)=\hat{\mathbf{u}}\cdot\mathbf{B}(x,y,h).
\end{equation}
For the contribution from one source pixel, this gives
\begin{equation}
    \Delta B_{\mathrm{NV}}=
    \frac{\mu_0 \Delta A}{4\pi r^3}
    \left[(-u_y h+u_z\Delta y)K_x+(u_x h-u_z\Delta x)K_y\right].
\end{equation}
Summing over all source pixels, the magnetic field contribution for pixel $i$ in the measurement plane is
\begin{equation}
    B_{\mathrm{NV}}^{(i)}=\sum_j G_{x,ij}^{(\hat{u})}K_x^{(j)}
              +\sum_j G_{y,ij}^{(\hat{u})}K_y^{(j)},
\end{equation}
with Green's matrices
\begin{equation}
    \begin{aligned}
        G_{x,ij}^{(\hat{u})}
        &=\frac{\mu_0\Delta A}{4\pi}
        \frac{-u_yh+u_z\Delta y_{ij}}
        {(\Delta x_{ij}^2+\Delta y_{ij}^2+h^2)^{3/2}},\\
        G_{y,ij}^{(\hat{u})}
        &=\frac{\mu_0\Delta A}{4\pi}
        \frac{u_xh-u_z\Delta x_{ij}}
        {(\Delta x_{ij}^2+\Delta y_{ij}^2+h^2)^{3/2}},
    \end{aligned}
    \label{eq:si_greens}
\end{equation}
where $\Delta x_{ij}=x_i-x_j$ and $\Delta y_{ij}=y_i-y_j$.
For the projection axis \(\hat{\mathbf{u}}\), both in-plane current components \(K_x\) and \(K_y\) therefore contribute to the measured projected field \(B_{\mathrm{NV}}\).

\subsection*{Differential magnetic-field measurements}

The projected magnetic field was sampled on a $100\times15$ grid covering $x\in[0,36]$ mm and $y\in[20,30]$ mm, with sampling intervals $\Delta x=0.36$ mm and $\Delta y=0.71$ mm. This gives 1500 measured field values. The scan was acquired line by line along $x$ while stepping the scan position in $y$. During movement in $x$, the sensor continuously sampled the field at 30.5 kHz, corresponding to the output sampling rate after lock-in decimation; the signal was already low-pass filtered by the 2 kHz lock-in bandwidth at this stage. The continuous data were binned into 100 pixels along $x$ according to stage position. Relative to the cell body, $x\in[3,31]$ mm, the scan extends slightly beyond the cell in $x$ but covers only a central strip in $y$. To isolate current-generated magnetic fields from the static field of the ferromagnetic anode tab and connectors, a no-current reference scan was subtracted from each charge and discharge map. Four scans were averaged for each operating direction. Mild Gaussian smoothing with $\sigma=(1.5,1)$ pixels was applied to suppress scan-to-scan noise from small stage offsets. The resulting measured field maps for charge and discharge, together with the reconstruction residuals are shown in Fig.~\ref{fig:si_residuals}.

\subsection*{Regularized real-space inversion}

The inverse problem is to find the current map underlying the measured magnetic-field data. A common solution is Fourier-space inversion, as often used for magnetic current imaging. For our measurements, however, that representation is not the most natural one. Our measured map is finite and strongly non-periodic: it covers the full cell width in $x$, but only a partial strip of the battery in $y$, so the currents extend beyond our measured $y$-window and still contribute to the measured field inside it. We therefore solve the problem in real space on a $1 \, \mathrm{mm}$ source grid restricted to $x \in [3,31] \, \mathrm{mm}$ and $y \in [17,33] \, \mathrm{mm}$. The measured strip $y \in [20,30] \, \mathrm{mm}$ is padded by $3 \, \mathrm{mm}$ on either side so that currents just outside the scan window can still contribute to the measured field. Outside this reconstruction region, the current density is set to zero.

The reconstructed current field $(\hat{\mathbf{k}}_x,\hat{\mathbf{k}}_y)$ is obtained by solving the following regularized least-squares problem on the source grid:
\begin{equation}
    \begin{aligned}
        (\hat{\mathbf{k}}_x,\hat{\mathbf{k}}_y)
        =\operatorname*{arg\,min}_{\mathbf{k}_x,\mathbf{k}_y}
        \Big[
        &\left\|\mathbf{G}_x^{(\hat{u})}\mathbf{k}_x+
        \mathbf{G}_y^{(\hat{u})}\mathbf{k}_y-\mathbf{b}\right\|_2^2\\
        &+\lambda\left(\|\mathbf{L}\mathbf{k}_x\|_2^2+
        \|\mathbf{L}\mathbf{k}_y\|_2^2\right)\\
        &+\mu_{\mathrm{div}}\|\mathbf{D}_x\mathbf{k}_x+
        \mathbf{D}_y\mathbf{k}_y\|_2^2\\
        &+\mu_{\mathrm{flux}}\|\mathbf{P}\mathbf{k}_y\|_2^2\\
        &+\mu_{\mathrm{amp}}\left(\|\mathbf{W}\mathbf{k}_x\|_2^2+
        \|\mathbf{W}\mathbf{k}_y\|_2^2\right)
        \Big].
    \end{aligned}
    \label{eq:si_objective}
\end{equation}
Here $\mathbf{b}$ is the measured field vector. The first term is the data-fidelity term: it minimizes the squared difference between the measured field and the field predicted by the forward model based on the reconstructed currents. The second term is a generalized Tikhonov smoothness penalty that suppresses current variations below the stand-off-limited spatial resolution. On the discrete source grid, it is the first-order finite-difference analogue of penalizing \(|\nabla K_x|^2+|\nabla K_y|^2\). The regularization parameter $\lambda$ controls the strength of this smoothness prior relative to the data-fidelity term and is selected by an L-curve method. The remaining terms encode additional physical and modeling priors. The third term penalizes nonzero current divergence, where \(\mathbf{D}_x\mathbf{k}_x+\mathbf{D}_y\mathbf{k}_y\) discretizes \(\partial_x K_x+\partial_y K_y\). The fourth term penalizes nonzero row-wise $y$-current flux. The operator \(\mathbf{P}\) acts on each source-grid row as
\[
    (\mathbf{P}\mathbf{k}_y)_m=\sum_n K_y(x_n,y_m),
\]
which is proportional to the slice-integrated flux \(\Phi(y_m)=\int K_y(x,y_m)\,\mathrm{d}x\). For the pouch-cell geometry used here, both tabs are located on the same edge, so the net $y$-directed current across interior horizontal slices is expected to vanish. The fifth term tapers current amplitudes in weakly constrained boundary regions. The weights in \(\mathbf{W}\) are zero inside the measured strip and increase smoothly as a half-cosine across the 3 mm extension bands to the outer reconstruction boundaries at \(y=17\) mm and \(y=33\) mm, suppressing solutions that place unrealistically large currents just outside the measured window. In addition, the implementation applies a separate cosine edge taper over 3 mm at the left and right cell boundaries in \(x\), suppressing boundary artefacts at these edges.

\subsection*{Parameter selection and normalization}

The parameters in Eq.~\eqref{eq:si_objective} are trace-normalized. In the normal equations solved numerically, the corresponding effective coefficients are
\begin{equation}
    \begin{aligned}
        \lambda^{\mathrm{eff}} &= \lambda \frac{T_G}{2T_L},
        &\mu_{\mathrm{div}}^{\mathrm{eff}} &= \mu_{\mathrm{div}}\frac{T_G}{T_{\mathrm{div}}},\\
        \mu_{\mathrm{flux}}^{\mathrm{eff}} &= \mu_{\mathrm{flux}}\frac{T_G}{T_{\mathrm{flux}}},
        &\mu_{\mathrm{amp}}^{\mathrm{eff}} &= \mu_{\mathrm{amp}}\frac{T_G}{T_{\mathrm{amp}}}.
    \end{aligned}
\end{equation}
Here \(T_G=\operatorname{tr}\!\left[(\mathbf{G}_x^{(\hat{u})})^T\mathbf{G}_x^{(\hat{u})}+(\mathbf{G}_y^{(\hat{u})})^T\mathbf{G}_y^{(\hat{u})}\right]\) is the trace of the data Hessian. The remaining traces are \(T_L=\operatorname{tr}(\mathbf{L}^T\mathbf{L})\), \(T_{\mathrm{div}}=\operatorname{tr}(\mathbf{D}_x^T\mathbf{D}_x+\mathbf{D}_y^T\mathbf{D}_y)\), \(T_{\mathrm{flux}}=\operatorname{tr}(\mathbf{P}^T\mathbf{P})\), and \(T_{\mathrm{amp}}=2\operatorname{tr}(\mathbf{W}^T\mathbf{W})\). With this convention, the quoted weights are dimensionless parameters that specify the relative strength of each prior with respect to the data term, rather than depending directly on units, grid spacing or matrix size.

After normalization, $\mu_{\mathrm{div}}=10^8$ and $\mu_{\mathrm{flux}}=10^6$ were used. These values suppress the residual divergence to approximately \(3\times10^{-5}\%\) of the local current per 1 mm source-grid cell for both charge and discharge reconstruction, and enforce the expected row-wise current balance without noticeably degrading the magnetic-field fit. The amplitude taper was set to $\mu_{\mathrm{amp}}=5$. If this parameter is too large, currents in the extension bands are forced artificially to zero; if it is too small, the reconstruction can place unphysically large currents in weakly constrained boundary regions.

\begin{figure}[htbp]
    \centering
    \includegraphics[width=0.72\linewidth]{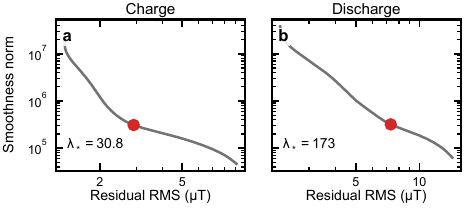}
    \caption{\textbf{L-curves for current-density reconstruction.} The selected marker denotes the maximum-curvature value $\lambda_*$ used for the reported charge and discharge reconstructions.}
    \label{fig:si_lcurves}
\end{figure}

The smoothness parameter $\lambda$ was selected separately for charge and discharge by sweeping 40 logarithmically spaced values from $10^{-6}$ to $10^4$ and selecting the maximum-curvature point of the L-curve \cite{si_hansen_analysis_1992} (Fig.~\ref{fig:si_lcurves}). The resulting L-curves have a broad bend rather than a sharply localized corner. In the L-curve framework, this indicates that a range of nearby regularization parameters gives a similar trade-off for the residual-versus-smoothness optimization that the L-curve method performs. In our data, this behavior is consistent with the measurement geometry: propagation over the 2 mm stand-off strongly attenuates fine-scale current variations, while the physical constraints already strongly reduce the remaining solution space on larger scales. We therefore use the maximum-curvature value as a parameter choice and interpret the reconstructed current maps only on the stand-off-limited length scale.

\subsection*{Interpretation and limitations}

The inferred current paths are consistent with current spreading from the cathode tab and converging toward the anode tab during charging, with the discharge case reversing sign. The reconstruction has several limitations. First, it yields only the depth-integrated sheet current, so individual electrode layers cannot be separated. Second, the approximately 2 mm stand-off suppresses fine spatial structure and limits practical resolution to a few millimeters. Third, only the strip $y\in[20,30]$ mm is directly constrained by measured data; the extension bands are included to capture edge contributions but should be interpreted cautiously. Additional uncertainty arises from imperfect knowledge of the NV-axis orientation and exact stand-off distance, both of which may vary slightly across the scan if the battery holder is tilted or rotated relative to the sensor plane. Finally, the effective-sheet model does not explicitly account for finite battery thickness.

\clearpage
\section*{Supplementary Note 9: Portable sensor backend}
\addcontentsline{toc}{section}{Supplementary Note 9: Portable sensor backend}

\begin{figure}[htbp]
    \centering
    \includegraphics[width=0.72\linewidth]{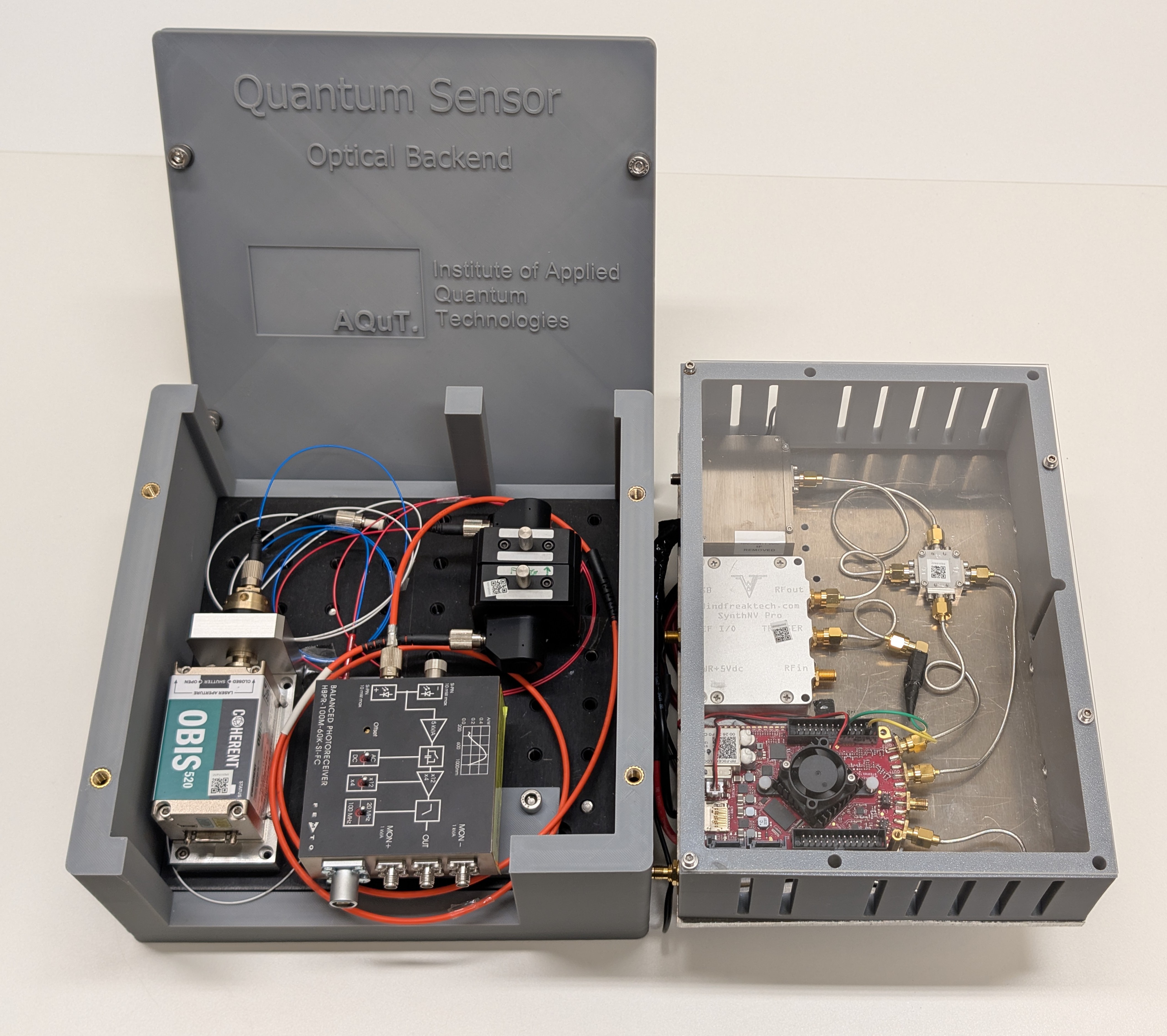}
    \caption{\textbf{Portable sensor backend.} The modular backend contains the optical and electronic components required for ODMR measurements, including a compact laser, photodetector, optical filter, and the FPGA-based control electronics. The optical and electronic components are mounted on aluminum plates respectively and housed in 3D-printed enclosures for mechanical stability and portability.}
    \label{fig:si_portable_backend}
\end{figure}

For portable operation of the sensor, we built a modular backend containing both the optical and electronic components required for ODMR measurements. The optical backend includes a compact laser (Coherent Obis 520 nm LX 40 mW), photodetector (Femto HBPR-100M-60K-SI-FC) and an optical filter. All components are fiber-connectorized. The electronic backend includes the same hardware used in the main text demonstrator, however substituting the amplifier with a more compact alternative (Minicircuits ZVE-3W-83+). The electronic components are mounted on a 8 mm thick aluminum plate for mechanical stability and heat dissipation, and the whole system is housed in a 3D-printed compact enclosure (Fig.~\ref{fig:si_portable_backend}). The modular design allows the sensor head to be easily connected and disconnected from the backend, enabling portable operation and flexible integration with different sensor heads or measurement setups. The compact form factor of the backend also facilitates transportation and deployment in various environments for field measurements or applications outside the laboratory setting.

\end{document}